# Boutiques: a flexible framework for automated application integration in computing platforms

Tristan Glatard[1], Gregory Kiar[2,3], Tristan Aumentado-Armstrong[2,3], Natacha Beck[2,3], Pierre Bellec[4], Rémi Bernard[2,3], Axel Bonnet[5], Sorina Camarasu-Pop[5], Frédéric Cervenansky[5], Samir Das[2,3], Rafael Ferreira da Silva[6], Guillaume Flandin[7], Pascal Girard[5], Krzysztof J. Gorgolewski[8], Charles R.G. Guttmann[9], Valérie Hayot-Sasson[1], Pierre-Olivier Quirion[4], Pierre Rioux[2,3], Marc-Étienne Rousseau[10] and Alan C. Evans[2,3]

[1]Department of Computer Science and Software Engineering, Concordia University, Montreal, Canada and [2]McGill University, Montreal, Canada and [3]Montreal Neurological Institute, Montreal, Canada and [4]Centre de Recherche de l'Institut de Gériatrie de Montréal CRIUGM, Montréal, QC, Canada and [5]University of Lyon, CNRS, INSERM, CREATIS, Villeurbanne, France and [6]University of Southern California, Information Sciences Institute, Marina del Rey, CA, USA and [7]Wellcome Trust Centre for Neuroimaging, London, UK and [8]Department of Psychology, Stanford University, Stanford, California, USA and [9]Center for Neurological Imaging, Department of Radiology, Brigham and Women's Hospital, Boston, Massachusetts, USA and [10]Compute Canada

## Abstract

We present Boutiques, a system to automatically publish, integrate and execute applications across computational platforms. Boutiques applications are installed through software containers described in a rich and flexible JSON language. A set of core tools facilitate the construction, validation, import, execution, and publishing of applications. Boutiques is currently supported by several distinct virtual research platforms, and it has been used to describe dozens of applications in the neuroinformatics domain. We expect Boutiques to improve the quality of application integration in computational platforms, to reduce redundancy of effort, to contribute to computational reproducibility, and to foster Open Science.

## Introduction

Computational platforms such as web services, portals, scientific gateways, workflow engines and virtual research environments commonly integrate third-party applications to enable various types of data processing. Applications, however, are often manually and repeatedly integrated whereas automating and sharing this effort would improve computational reproducibility [1, 2] and contribute to Open Science. Meanwhile, container systems such as Docker[1] and Singularity [3] have emerged to facilitate the sharing and migration of software by defining immutable, reusable execution environments.

We present Boutiques, a system to publish, integrate and execute command-line applications across platforms (see Figure 1). In Boutiques, a command line is described using a flexible template comprising the inputs it requires and the outputs it produces. Inputs may be passed directly on the command line or through configuration files. They may also be interdependent, for instance mutually exclusive. Such formal descriptions, simply referred to as *descriptors*, link to a container image where the given application is installed. Boutiques descriptors allow for automatic application integration in platforms, and advanced validation of input values to prevent errors. Boutiques descriptors are intended to be produced by scientific application developers, stored alongside their application, indexed by common repositories, and consumed by execution platforms. A set of core tools facilitate the construction, validation, import, execution, and publishing of Boutiques descriptors.

The remainder of this paper describes the Boutiques system and reports on its adoption by platforms and applications in the neuroinformatics domain, our primary field of interest. It closes on a discussion and comparison with related systems.

---

[1] https://www.docker.com

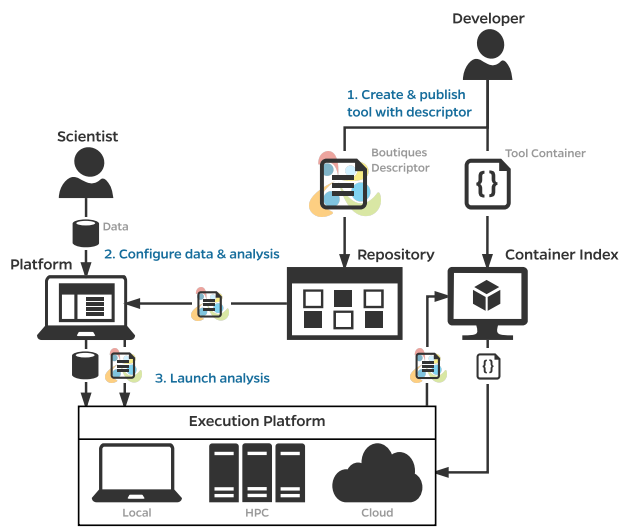

**Figure 1.** Publication, integration and execution of applications with Boutiques.

## System description

In Boutiques, applications are described with a JSON descriptor that specifies the command-line template, inputs and outputs. The descriptor may point to a container where the application and all its dependencies are installed. It may also contain an invocation schema used for input validation (this will be created at runtime if it is not found). At runtime, the execution platform builds the command line from the descriptor and the values entered by the user. The platform runs the command line on the execution infrastructure, e.g., a server, a cluster or a cloud, within a container whenever available. To build and run the command line, the platform may rely on the Boutiques core tools, in particular the validator and executor, packaged through the `bosh` command-line utility.

## Command-line description

The core component of the descriptor is a command-line template complying to the syntax of the `bash` UNIX shell, the default shell on most of the Linux distributions and on OS X. The command-line template is a single string that may contain placeholders for input and output values, called value keys. It may also encompass several commands separated by `bash` constructs such as semicolons, pipes (`|`) or ampersands (`&`), to facilitate the embedding of basic operations on the command line, for instance directory creation, input decompression or output archival.

Here is an example of a typical command-line template:

```
exampleTool_1 [CONFIG_FILE] [STRING_INPUT] [FILE_INPUT] | \
  exampleTool_2 [FLAG_INPUT] [NUMBER_INPUT] >> [LOG].txt
```

The template contains five value keys, identified by square brackets, that will be replaced by values and file names according to the user input when the application is executed. Flags will also be added wherever appropriate, with customizable separators. Value keys have to be unique but do not have to comply to any particular syntax. Note the use of the `|` operator to chain applications, and of the `»` operator to redirect the standard output to a file.

## Input description

*General properties.* Inputs must have a name, a unique identifier, and a type. They may be optional, have a description, a value key, a flag and flag separator, and a default value. Inputs may also be ordered lists: in this case, value keys are substituted by the space-separated list of input values.

*Types.* Inputs may be of type `String`, `Number`, `Flag` or `File`. `File` may also represent a directory. Types can be restricted to a specific set of values or to a specific range.

*Groups and dependencies.* Groups of inputs may be defined with an identifier, name, and list of input identifiers. Groups may be used to improve the presentation in a graphical user interface, and to specify the following constraints among inputs: (1) `mutually-exclusive`: only one member in the group may have a value; (2) `one-is-required`: at least one member in the group must have a value; (3) `all-or-none`: if any of the members have a value then all members must have a value. Dependencies among inputs may also be defined regardless of a particular group: an input may (1) require a list of inputs and (2) disable a list of inputs.

Listing 1 shows the definition of an input in the command line exemplified above. According to this definition and assuming that the input value entered by the user is 0.3, the string `[NUMBER_INPUT]` will be replaced by `-n=0.3` on the command line.

```
{
    "id" : "num_input",
    "name" : "A number input",
    "type" : "Number",
    "value-key" : "[NUMBER_INPUT]",
    "optional" : true,
    "command-line-flag" : "-n",
    "command-line-separator" : "=",
    "minimum" : 0,
    "maximum" : 1,
    "exclusive-minimum" : true,
    "exclusive-maximum" : false
}
```

**Listing 1.** Example of a `Number`-type input.

## Output description

Application outputs are files and directories that need to be delivered to the user once the execution is complete. Outputs need to be specified so that computing platforms can identify the files that must be saved after the execution and raise errors if they are not present.

In Boutiques, output files must have a unique identifier, a name, and a path template that specifies the file or directory name. Path templates may include input value keys in case output files are named after the input values. In this case, input values may be stripped from specific strings, e.g., file extensions, before being substituted in the path template. Output files may also have a description, a command-line flag, a flag separator, and a value key in case they appear on the command line. They may be optional in case the file is not always produced by the application, for instance when it is produced only when a particular flag is activated. They may also be lists: in this case, the path template must contain a wildcard (*) matching any string of characters and defining the pattern used to match the output files in the list.

Listing 2 shows the definition of an output file in the command line exemplified before. According to this definition and assuming that the string input value entered by the user is `foo.csv`, the string `[LOG]` will be replaced by `log-foo` on the command line.

```
{
  "id": "logfile",
  "name": "Log file",
  "description": "The output log file",
  "path-template": "log-[STRING_INPUT]",
  "value-key": "[LOG]",
  "path-template-stripped-extensions": [".txt", ".csv"],
  "optional": false
}
```

**Listing 2.** Example of an output leveraging `path-template` search-and-replacement.

## Configuration files

A large number of applications rely on configuration files rather than command-line options to define their input and output parameters. As the number of parameters increases, command lines rapidly become long and cumbersome whereas configuration files allow for better structure and documentation.

Configuration files may be complex though, and specified in any language. For this reason, Boutiques allows application developers to specify their own template containing input and output value keys. Configuration files are specific types of output files that must have a file template that defines how they will be named and where they will be written. They may also have a value key and a flag in case they need to be passed on the command line. Listing 3 shows an example.

```
{
    "id": "config_file",
    "name": "Configuration file",
    "type": "File",
    "value-key": "[CONFIG_FILE]",
    "path-template": "config.txt",
    "file-template": [
        "# This input is hard-coded",
        "stringInput=foo",
        "# This is an input file",
        "fileInput=[FILE-INPUT]",
        "# And here is the result",
        "fileOutput=[OUTPUT-FILE-NAME]",
        ""
    ]
}
```

**Listing 3.** Example of a configuration input file. The file template is defined as an array of strings to allow for multi-line strings in JSON.

## Command-line construction

At runtime, a value is assigned to all the mandatory and some of the optional inputs. Algorithm 1 shows how the command line is constructed from the descriptor and the values entered. It substitutes all the value keys in the command line, output path templates and configuration files, and writes the configuration files.

---

**Algorithm 1** Command-line construction

\# *Substitute input value keys in output path templates, configuration files and command line.*
**for** input in inputs **do**
   **if** input has a `value-key` **then**
      **for** output in outputs **do**
         stripped_value = input_value
         **if** input type is File **then**
            In stripped_value, remove all elements in `path-template-stripped-extensions`.
         **end if**
         In `path-template`, replace all occurrences of `value-key` by stripped_value.
         **if** output has a `file-template` **then**
            In any line of `file-template`, replace all occurrences of `value-key` by stripped_value.
         **end if**
      **end for**
      Prepend `command-line-flag` and `command-line-flag-separator` to input_value.
      In `command-line`, replace all occurrences of `value-key` by input_value.
   **end if**
**end for**
\# *Substitute output value keys in configuration files and command line.*
**for** output in outputs **do**
   **if** output has a `value-key` **then**
      **for** output in outputs **do**
         **if** output has a `file-template` **then**
            In any line of `file-template`, replace all occurrences of `value-key` by `path-template`.
            \# *Input `value-key` have been substituted in `path-template` previously.*
         **end if**
      **end for**
      Prepend `command-line-flag` and `command-line-flag-separator` to `path-template`.
      In `command-line`, replace all occurrences of `value-key` by `path-template`.
   **end if**
**end for**
\# *Write all configuration files.*
**for** output in outputs **do**
   **if** output has a `file-template` **then**
      Write `file-template` in `path-template`
      \# *Value keys have been substituted in `file-template` previously.*
   **end if**
**end for**

---

## Invocation schema

Rigorous input validation is an important motivation for Boutiques. For this purpose, Boutiques relies on an application-specific JSON schema, called *invocation schema*, to specify the input values accepted by an application. Platforms can rely on invocation schemas to validate inputs using any JSON validator, without having to develop specific code.

Invocation schemas, however, are complex JSON objects. Basically, they must represent the properties described above in a formal way, including dependencies between inputs. Listing 4 shows an example of how dependencies between mutually exclusive parameters are defined in the invocation schema. To relieve application developers from the burden of having to write JSON schemas, invocation schemas can be generated automatically by the `bosh` command-line utility. The invocation schema is stored as an optional property of the Boutiques descriptor.

```
{
  "dependencies" : {
     "num_input" : {
        "properties" : {
           "str_input" : {
              "not" : {}
           }
        }
     },
     "str_input" : {
        "properties" : {
           "num_input" : {
              "not" : {}
           }
        }
     }
  }
}
```

**Listing 4.** Excerpt from invocation schema showing dependencies between two mutually exclusive parameters `num_input` and `str_input`.

### Workflow support

Boutiques does not specify a particular language to build workflows from descriptors, due to the large amount of specialized frameworks to do so. However, workflows can be both composed from and described as Boutiques descriptors: workflow engines can leverage the `bosh` tools to call Boutiques applications from their descriptors; in turn, workflows can be described as Boutiques descriptors. Such a "task encapsulation" model allows for a scalable and reliable execution of workflows expressed in a variety of languages, as detailed in [4].

### Containers

Applications may be installed in a container image complying to the Docker, Singularity or rootfs format. We intentionally support multiple container formats as we anticipate that they will be used for different purposes. For instance, Docker is well suited for developers and users who manipulate applications on their local workstations or the cloud. It is well documented, maintained and it has a rich ecosystem of tools to build and run containers on most operating systems. Singularity is more suited for users and platforms that need to run applications on shared computing clusters. Bridges exist among these containers formats to convert container images across frameworks. For instance, a platform dedicated to high-performance computing may accept descriptors referring to Docker containers to facilitate application integration by developers, and run these images on clusters using Singularity.

Container images are defined from their URL (rootfs) or image name in a Docker or Singularity index. Descriptors may specify a working directory where the application has to be run, they may indicate if the image has an entry point, and they may also report a hash to accurately identify container images and detect updates.

Containers were adopted because they allow for an automated and lightweight integration of application implementations in platforms. They are extremely useful to improve the reproducibility of analyses as variations in the software environment may have an important impact on the computed results. They also have limitations, in particular they do not specify the hardware architecture required to execute an application, which can be an issue in some cases.

### Resource requirements

Boutiques descriptors may contain requirements regarding the number of CPU cores or nodes, the amount of RAM or disk storage, and the total walltime expected for a typical execution of the application. Such properties are called "suggested resources" as we are well aware that the actual resource requirements usually depend on the input data, parameters and hardware infrastructure.

### Custom properties

Custom properties may be added to the Boutiques specification without restriction. Custom properties are grouped together in a specific JSON object to facilitate validation. They may be useful to implement platform-specific features but they should be used with care to avoid making applications dependent on a particular platform or replacing existing functionality already represented in the Boutiques schema.

### Core tools

Boutiques is available on Python through the PyPi package repository as "`boutiques`". The Boutiques package exposes a command-line utility to the user, `bosh`, which contains entry-points for all core functions within Boutiques. The core tools provided by Boutiques are, the: validator, executor (for both launch and simulation), invocation schema handler, importer, and publisher. The tools exposed through the command-line interface are also available consistently through a Python API by importing the Boutiques package. Though not a component of the Boutiques tool chain, a Jupyter Notebook tutorial exists to facilitate new users getting started with Boutiques.

*Validator.* The Boutiques validator checks conformance of JSON descriptors to the Boutiques schema using a basic JSON validator. It also performs the following checks that cannot be easily implemented in JSON schema: value keys are unique among inputs, input and output identifiers are unique, input and output value keys are all included in the command line, identifiers with the same value key are mutually exclusive, value keys are not contained within each other (which would puzzle substitution), output path templates are unique (to avoid results to overwrite each other), inputs of type Flag have a command-line flag, are optional and are not lists, the default value of restricted types is part of the restriction, an input cannot both require and disable another input, required inputs cannot require or disable other parameters, group member identifiers must correspond to existing inputs and cannot appear in different groups, mutually exclusive groups cannot have members requiring other members, one-is-required groups should never have required members, and all-or-none group members must not be required.

*Executor.* The executor has two modes of operation: *simulate* and *launch*. The simulate mode can generate hypothetical command lines from random values given the descriptor (and corresponding invocation schema) for debugging purposes, or display the command that would be executed given a provided valid invocation. The launch mode can execute command line from a Boutiques descriptor and a set of input values represented in JSON file complying with the invocation schema. It runs the command in a container provided that the required framework (e.g. Docker) is installed. The executor can be used by application users to run applications locally, or by platforms to generate command lines to be run on the execution infrastructure.

*Invocation Schema Handler.* The invocation schema handler can create an invocation schema from a Boutiques descriptor and validate input data against it using a regular JSON validator. It can be used to add invocation schemas to existing descriptors. It is used by the executor if no invocation schema is present in the Boutiques descriptor being deployed.

*Importer.* The importer takes Boutiques descriptors from older versions and updates them to be compliant with the most recent version of the schema. This tool can also create descriptors from selected application collections, such as BIDS apps [5].

*Publisher.* As Boutiques has primarily been adopted in the neuroinformatics community, the publisher gets a further description (such as author, website, etc.) of the described application, and adds an index to it on NeuroLinks[2], a repository containing links to neuroinformatics resources and tools. This functionality could be extended to new repositories for other domains, such as Bioconductor for bioinformatics [6].

# Results

## Supported platforms

The import and execution of Boutiques applications is currently supported in the platforms enumerated below.

### CBRAIN
CBRAIN (http://github.com/aces/cbrain) [7] is a web platform to process data distributed into multiple storage locations on computing clusters and clouds. CBRAIN offers transparent access to remote data sources, distributed computing sites, and an array of processing and visualization tools within a controlled, secure environment. The CBRAIN service deployed at the Montreal Neurological Institute relies on the infrastructure provided by Compute Canada [8]. It currently provides 500+ collaborators in 22 countries with web access to several systems, including 6 clusters of the Compute Canada high-performance computing infrastructure (totaling more than 100,000 computing cores and 40 PB of disk storage) and Amazon EC2. CBRAIN transiently stores about 10 million files representing over 50 TB distributed in 42 servers. 51 public data processing applications are integrated and over 340,000 processing batches have been submitted since 2010.

Applications in CBRAIN are integrated as Ruby classes that create web forms, validate parameters and run command lines on computing resources. Boutiques is supported through a set of templates that generate such classes from the application descriptor. Two application integration modes are available:

i. The descriptor is stored in a CBRAIN plugin and the Ruby classes are generated on-the-fly when CBRAIN starts. This mode allows CBRAIN developers to update all Boutiques applications at once by editing the templates. However, it does not allow for customization beyond the Boutiques schema. To provide more flexibility, we added a custom property (`cbrain:inherits-from-class`) to the Boutiques descriptor to define the Ruby class that should be used as parent class for the application.
ii. Ruby classes are generated from descriptors through an offline process and integrated in CBRAIN as any other application. This mode allows developers to customize applications by editing the generated Ruby classes, but the resulting applications are difficult to maintain in the long term, in particular when the descriptors are updated.

We also extended CBRAIN to enable the parallelization of workflows wrapped as Boutiques descriptors. Applications with the `cbrain:can-submit-new-tasks` custom property may submit sub-tasks by creating Boutiques invocations in their working directory. CBRAIN periodically scans working directories, submits the requested sub-tasks, and writes back an invocation identifier in the same directory. This parallelization model is simple, the application only needs to write Boutiques invocations and communication happens through the file system, and it is also powerful as it enables the parallelization of complex workflows such as the Niak ones described later.

We also introduced a new list mechanism in CBRAIN to facilitate the iteration of Boutiques applications on large sets of files. CBRAIN lists are specific files that contain references to other CBRAIN files. When a list is passed to a Boutiques application, the elements in the list are either concatenated in a single command line (when the corresponding Boutiques input is a list), or a new command line is generated for every element in the list (when the input is not a list). Supporting lists as a specific CBRAIN file type allows for improved validation. For instance, lists that contain references to non-existent or deleted files can be detected. It also allows users to edit lists using their own tools such as scripts or spreadsheet applications.

### Nipype
Nipype (http://nipype.readthedocs.io/en/latest) [9] is a workflow engine widely used in neuroinformatics. Nipype workflows can be composed from Boutiques applications using the Python API. As an example, we implemented NipBIDS[3], a Nipype workflow to process BIDS datasets using BIDS apps imported as Boutiques applications. NipBIDS iterates participant analyses on all the subjects found in a BIDS dataset and runs a group analysis if requested.

### SPINE
SPINE (http://spinevirtuallab.org), which stands for Structured Planning and Implementation of New Explorations, is a web-based, collaborative platform (virtual laboratory) designed to support the design and execution of experiments centered on specific scientific questions. SPINE enables dis-tributed data collection and management, as well as experiment design, execution, and review. Boutiques will serve as SPINE's algorithm and workflow repository, and enables unequivocal referencing of specific workflows applied to specified datasets within an experiment, thereby describing the provenance and facilitating the reproducibility of image-derived measurements. Workflows in SPINE may combine human image annotation with automated image processing algorithms. Future development will focus on extending Boutiques workflow descriptors to include the identification and characterization of human operators and their specific historic performance on the required tasks. SPINE is currently hosted at Brigham & Women's Hospital in Boston, and supports several international projects.

---

2 https://brainhack101.github.io/neurolinks

3 https://github.com/big-data-lab-team/sim/tree/master/sim/other_wf_examples/nipype

*VIP*
The Virtual Imaging Platform (VIP) [10] is a web portal for medical simulation and image data analysis. VIP makes applications available as services, and connects them to the biomed Virtual Organization (VO) in the European Grid Infrastructure[4]. The biomed VO interconnects approximately 65 computing sites world-wide and provides access to 130 computing clusters and 5 PB of storage. The VIP service is deployed at the Creatis laboratory[5] in Lyon and it uses the DIRAC French national service[6] to execute jobs on EGI grid and cloud resources. As of October 2017, VIP counts more than 1,000 registered users and a growing number of available applications.

Until recently, applications were manually integrated in VIP as workflows written in the Gwendia [11] language and executed with the MOTEUR [12] engine. As of today, Boutiques is supported through an importer tool that parses the JSON descriptor and automatically generates the corresponding application workflow and the wrapper script that handles, among other things, the execution of the command line. In VIP, application workflows enable (1) iterations on input lists, (2) the generation of parallel tasks, and (3) the concatenation of multiple applications. For example, the applications used in the MICCAI challenges described below required workflows to evaluate results using the metrics defined by the challenge. Application concatenation is handled at the importer level based on pre-defined workflow templates.

### Integrated applications

Dozens of neuroinformatics applications were integrated in CBRAIN or VIP using Boutiques. The main ones are described below. Several Boutiques descriptors were published in Neurolinks through the `bosh` publisher.

*Anatomical imaging*
*FSL.* Several MRI analysis applications from the FMRIB Software library (FSL [13]) were integrated in CBRAIN using Boutiques: BET, fsl_anat, FAST and FIRST. Descriptors are on Neurolinks.

*Anima.* `animaN4BiasCorrection`, an ITK-based bias field correction application from the Anima[7] project was made available in VIP through Boutiques.

*Functional MRI (fMRI)*
*Niak.* The Niak fMRI pre-processing pipeline [14], executed with the Pipeline System for Octave and Matlab (PSOM) [15], was integrated in CBRAIN through Boutiques. The integration uses the CBRAIN sub-tasking mechanism described earlier so that even the invocations processing a single subject can be parallelized. It also allows CBRAIN to leverage the efficient agent model used in PSOM, as described in [16]. The integration required some work in PSOM to facilitate its invocation as a non-interactive command-line application. The resulting CBRAIN plugin is available at https://github.com/SIMEXP/cbrain-plugins-psom. Descriptors are on Neurolinks.

*GinFizz.* We integrated the GinFizz[8] Nipype-based fMRI preprocessing pipeline in VIP with Boutiques. A few technical issues coming from the management of users in Docker containers had to be addressed: to enable the execution in Boutiques we had to override the permissions of files and folders in the GinFizz container. We also had to install all the GinFizz pipeline components in a single container while multiple ones were used by the application initially.

*Diffusion imaging*
*MRTrix3.* A few applications from the MRtrix3 package [17][9] for diffusion MRI processing were also made available in VIP, as well as a pipeline developed at Creatis, which combines MRtrix3 and FSL applications.

*ndmg.* The NeuroData MRI to Graphs one-click connectome estimation pipeline [18], developed in Python and leveraging FSL, was exported to Boutiques and is available at https://github.com/neurodata/boutiques-tools. The ndmg pipeline is deployed in CBRAIN via its Boutiques descriptor, and is available both through Docker and Singularity container environments.

*Image simulation*
*CreaPhase.* The CreaPhase phase-contrast simulator, developed at Creatis, was integrated in VIP through Boutiques. The inputs had certain particularities (some needed to be enclosed in simple quotes, other were vectors of variable size enclosed in brackets) that required the post-processing of the wrapper script generated by the Boutiques importer.

*ODIN.* The Odin MRI simulator [19][10] was integrated in VIP through Boutiques. Since Odin requires important amounts of computing resources it is executed by VIP on the EGI grid that currently does not support Docker. The Docker image was used just for compilation and testing; the Odin executable was extracted from the Docker image and the Odin wrapper script was modified accordingly.

*BIDS apps*
BIDS apps [5], an effort for the adoption of the Brain Imaging Data Structure (BIDS) in common neuroimaging pipelines, require a standardized set of input and output parameters. We developed a tool as part of the Boutiques importer to generate a descriptor for any such BIDS app. We validated this tool by importing BIDS apps containing the Statistical Parametric Mapping toolbox (SPM) [20] and the ndmg pipeline mentioned above. Descriptors are on Neurolinks.

*2016 MICCAI challenges*
We used Boutiques to integrate 23 pipelines in the VIP platform in the context of two challenges organized by the MICCAI conference in 2016, related to the segmentation of multiple-sclerosis lesions in MR images (MSSEG challenge[11]) and of tumor volumes in PET images (PETSEG challenge [21]). The pipelines were integrated in VIP and executed on 205 subjects in a few weeks only. Some pipelines had to be adjusted manually once integrated in the platform for the following reasons:

i. A pipeline required a GPU, which we enabled through the nvidia-docker[12] tool not supported in Boutiques although it could be a possible extension.
ii. A pipeline required more than 10 GB of data dependencies (atlas data) which exceeded the maximum size allowed for Docker containers in our setup. We solved the issue by installing the data in a directory of the host server that we mounted in the container.
iii. A pipeline wrote more than 10 GB of intermediate data in a temporary directory of the container located on a 2 GB partition. We solved the issue by mounting a host directory in the temporary directory.

---

[4] https://www.egi.eu
[5] https://www.creatis.insa-lyon.fr
[6] https://dirac.france-grilles.fr
[7] https://github.com/Inria-Visages/Anima-Public/wiki
[8] https://github.com/thomashirsch/ginfizz
[9] http://www.mrtrix.org/
[10] http://od1n.sourceforge.net

Two custom properties (vip:miccai-challenger-email and vip:miccai-challenge-team-name) were also added to the Boutiques descriptor to help post-process results in the specific context of MICCAI challenges.

## Discussion

With Boutiques, developers can integrate their applications once and execute them in several platforms. Boutiques removes the technological dependency to a particular platform and facilitates application migration. Although the motivating use cases were taken from neuroinformatics, our primary field of interest, nothing prevents the system from being used in other domains.

### Boutiques descriptor

The Boutiques descriptor specification allows describing a wide range of applications, but it is also getting increasingly complex through additions such as invocation schemas and dependencies among inputs. Extending the Boutiques descriptor has two main goals: (1) validation: incorrect input values and execution results are more precisely detected when the application descriptor is comprehensive; (2) automation: a rich descriptor schema reduces the need for custom application wrappers, which is particularly useful for containerized applications.

Nonetheless, a complex descriptor schema has a cost for application developers and platforms, which we address as follows. For developers, we maintain the set of mandatory descriptor properties as small as possible so that simple applications can be described in a few lines only (see Listing 5). For platforms, we aim at supporting as many features as possible in the bosh executor so that only the following steps need to be implemented in a platform, regardless of the complexity of the descriptor:

i. Input entry: generate the interface to enter inputs.
ii. Input validation: create a JSON invocation from the interface, validate it against the invocation schema.
iii. Input delivery: transfer the input files to the application execution location.
iv. Execution: pass the invocation to the bosh executor to run the application.
v. Output delivery: from the descriptor, identify the output files and deliver them to the user.

In particular, command-line generation and advanced validation features such as dependencies between inputs are embedded in bosh, without requiring the platform to support the related descriptor properties.

```
{
  "name": "echo",
  "tool-version": "1.0",
  "description": "A simple script to test output files",
  "command-line": "echo [PARAM] > output.txt",
  "schema-version": "0.5",
  "inputs": [{
      "id": "param",
      "name": "Parameter",
      "value-key": "[PARAM]",
      "type": "Number"
  }],
  "output-files": [{
      "id": "output_file",
      "name": "Output file",
      "path-template": "output.txt"
  }]
}
```

**Listing 5.** A minimal Boutiques descriptor.

### Workflow support

Boutiques intentionally does not provide a workflow language to compose applications, as this is already possible with numerous workflow engines. Workflows can be either composed from Boutiques applications, as we illustrated with the Nipype and MOTEUR engines, or described as Boutiques applications, as we demonstrated with the Niak fMRI pre-processing pipeline. Furthermore, the CBRAIN platform has a sub-tasking mechanism that allows Boutiques applications to submit tasks, which is used to parallelize workflows wrapped as Boutiques applications. Based on the CBRAIN experience, we may specify the sub-tasking mechanism in Boutiques so that other platforms can benefit from it. This model is powerful because it shields the Boutiques specification from specific workflow constructs and it allows a wide range of workflow engines to be described and used uniformly.

### Reproducibility

Boutiques helps computational reproducibility through containers and formal command-line descriptions. With Boutiques, complete sets of applications could be easily migrated across execution platforms, including high-performance computing clusters and individual laptops, to reproduce analyses. The Boutiques descriptor describes the parameters and implementation of the application, and the invocation schema describes the parameter values.

However, reproducibility is a large problem that Boutiques only partially addresses. At the command-line execution level, containers help freeze a large fraction of the software ecosystem but they do not shield against discrepancies arising from different Linux kernel versions or hardware platforms. For instance, containers may not execute consistently on differ-ent CPUs complying to the x86_64 architecture (e.g. Intel and AMD) when the application is compiled with architecture-specific flags such as GCC's `-march`.

---

[11] https://portal.fli-iam.irisa.fr/msseg-challenge/overview
[12] https://github.com/NVIDIA/nvidia-docker

In addition, important runtime parameters, for instance related to multi-threading or available resources (storage, RAM), may be set by the execution platform without being specified in the Boutiques descriptor. Such runtime parameters may impact reproducibility in some cases. To properly cover this issue, Boutiques descriptors should be complemented by a provenance framework that captures a detailed trace of the execution. We plan to leverage the provenance format being defined by the NeuroImaging Data Model-Workflow [22] initiative for this purpose.

### Application types

So far, Boutiques has focused on the description of non-interactive command-line applications. While such applications cover a large fraction of the applications involved in scientific data processing, other types of programs exist such as web services, interactive applications, and applications with a graphical user interface (GUI). Such application classes could be described in Boutiques through a command-line mapping. For instance, web services may be wrapped as command-line applications using tools such as `curl` or `wget`. Interactive applications may also be transformed to non-interactive ones through configuration files. Finally, nothing prevents a Boutiques application from popping up a window for a user to provide input through a GUI. This should, however, be specified as an extension to the descriptor since most platforms would not support this feature by default. Graphical output produced by applications executed in containers should also be treated specifically.

### Limitations

A few limitations remain that should be addressed in the future. First, Boutiques moves the application integration bottleneck from integration to validation. Using Boutiques, functions can be automatically exported from frameworks such as Nipype and SPM, creating hundreds of richly described applications potentially usable by end-users. However, the automated validation of such applications remains challenging. A Boutiques-specific testing framework could be designed and potentially fed by existing frameworks, to address this issue.

Another limitation is related to the security of containerized applications. Since containers are usually controlled by application developers rather than platforms, which is a good thing to reduce application integration bottlenecks, nothing prevents developers to embed malicious code in their container at any stage of the process, possibly after a platform administrator inspected the container. Containers are bulky file archives that are cumbersome to inspect. Tools need to be developed to allow for an easier characterization of container contents, for instance through comparison digests with respect to validated base images. Singularity containers have reduced security risks as compared to Docker containers, but the issue of content transparency is still not avoided.

### Related work

Several frameworks have been developed to describe and integrate applications in various types of platforms. Boutiques focuses on (1) fully-automatic integration of applications, including deployment on heterogeneous computing resources through containers, (2) comprehensive input validation through a strict JSON schema, (3) flexible application description through a rich JSON schema.

### Common Workflow Language

The Common Workflow Language (CWL[13]) is the work most closely related to Boutiques as it provides a formal way to describe containerized applications. In particular, CWL's Command Line Tool Description overlaps with the Boutiques descriptor. This Section highlights the main differences between CWL and Boutiques, based on version 1.0 of the CWL Command Line Tool Description[14]. According to GitHub, CWL started 6 months before Boutiques (September 2014 vs. May 2015).

*Conceptual differences*
The following differences are conceptual in the sense that they may not be easily addressed in CWL or Boutiques without deeply refactoring the frameworks.

First, CWL has a workflow language whereas Boutiques does not. In Boutiques, workflows are integrated as any other applications, except that they may submit other invocations to enable workflow parallelism. This fundamental difference has consequences on the complexity of CWL application descriptions and on the possibility to reuse existing workflows in Boutiques. The adoption of ontologies in CWL may also be another consequence (see below).

CWL imposes a strict command-line format while Boutiques is more flexible. CWL specifies command lines using an array containing an executable and a set of arguments whereas Boutiques only uses a string template. Boutiques' template approach may create issues in some cases, but it also allows developers to add simple operations to an application without having to write a specific wrapper. For instance, a Boutiques command line may easily include input decompression using the `tar` command in addition to the main application command. Importantly, Boutiques' template system allows supporting configuration files.

CWL uses ontologies while Boutiques does not. Ontologies allow for richer definitions but they also have an overhead. The main consequences are the following:

- CWL uses a specific framework for validation, called SALAD (Semantic Annotations for Linked Avro Data) whereas Boutiques uses plain JSON schema. The main goal of SALAD is to allow "working with complex data structures and document formats, such as schemas, object references, and namespaces". Boutiques only relies on the basic types required to describe and validate a command line syntactically. While the use of SALAD certainly allows for higher-level validation, and may simplify the composition and validation of complex workflows, it also introduces a substantial overhead in the specification, and platforms have to use the validator provided by CWL. On the contrary, a regular JSON validator can be used in Boutiques.
- CWL has a rich set of types whereas Boutiques only has simple types. This may again be seen as a feature or as an overhead depending on the context. Boutiques tries to limit the complexity of the specification to facilitate its support by platforms where applications will be integrated.

*Major differences*
The following differences are major but they may be addressed by the CWL and Boutiques developers as they do not undermine the application description model.

---

13 http://www.commonwl.org
14 http://www.commonwl.org/v1.0/CommandLineTool.html

- CWL applications have to write in a specific set of directories called "designated output directory", "designated temporary directory" and "system temporary directory". Applications are informed of the location of such directories through environment variables. Having to write in specific directories is problematic because applications have to be modified to enable that. In Boutiques, the path of output files is defined using a dedicated property.
- CWL types are richer, not only semantically but also syntactically. For instance, files have properties for basename, dirname, location, path, checksum, etc.
- Boutiques supports various types of containers (Docker, Singularity, rootfs) while CWL supports only Docker. Both tools have rich requirements: for instance, they may include RAM, disk usage, and walltime estimate. CWL has hints, i.e., recommendations that only lead to warnings when not respected.
- In Boutiques, dependencies can be defined among inputs, e.g., to specify that an input may be used only when a particular flag is activated. This is a very useful feature to improve validation, in particular for applications with a lot of options.
- In Boutiques, named groups of inputs can be defined, which improves the presentation of long parameter lists for the user and enables the definition of more constraints within groups (e.g. mutually exclusive inputs).

### BIDS apps

BIDS apps [5] specify a framework for neuroimaging applications to process datasets complying to the Brain Imaging Data Structure (BIDS). They share common goals with Boutiques, in particular reusability across platforms through containerization. Conceptually, however, BIDS apps and Boutiques are different since BIDS apps intend to standardize application interfaces while Boutiques intends to describe them as flexibly as possible. BIDS apps have a specific set of inputs and outputs, for instance the input dataset, that have to be present in a specific order on the command line for the application to be valid. The specification adopted by BIDS apps simplifies the integration of applications in platforms as they all comply to the same interface. However, it is also limited to the subset of neuroimaging applications that process BIDS datasets and it does not formally describe application-specific inputs. All in all, BIDS apps and Boutiques complement each other: BIDS apps provide a practical way to integrate neuroimaging applications, while Boutiques offers a formal description of their specific parameters. Boutiques descriptors can be generated from BIDS apps using the `bosh` importer.

### Other frameworks

Several other frameworks have been created to facilitate the integration of command-line applications in platforms. In neuroinformatics, many platforms define a formal interface to embed command-line applications. Among them, the Common Toolkit[15] interoperates with several platforms such as 3D Slicer [23], NiftyView [24], GIMIAS [25], MedInria [25], MeVisLab [26] and MITK workbench [27]. The framework, however, remains tightly bound to the Common Toolkit's C++ implementation which limits its adoption, e.g., in web platforms.

In the distributed computing community, systems were also proposed to facilitate the embedding of applications in platforms. The Grid Execution Management for Legacy Code Architecture (GEMLCA [28]) was used to wrap applications in grid computing systems. Interestingly, it has been used to embed workflow engines in the SHIWA platform [29], in a similar but different way than proposed by Boutiques.

The recent advent of software containers requires a new generation of application description frameworks that are independent from any programming language and that expose a rich set of properties to describe command lines, as intended by Boutiques.

### Conclusion

Boutiques is available at `https://github.com/boutiques`. We welcome feedback, issue reporting and pull requests. Boutiques adopts a bottom-up approach where new features are progressively added based on feedback from applications and platforms. Beyond the technicalities discussed in this paper, the availability of a solid core of applications and platforms in the framework is key to its success, which we plan to continuously enhance.

### Acknowledgments

Pipeline integration for the MICCAI 2016 challenge was funded by the French National Agency for Research (ANR) through "France Life-Imaging". We also thank Compute Canada and Calcul Québec for providing a computing infrastructure supporting Docker and Singularity containers. This research was undertaken thanks in part to funding from the Canada First Research Excellence Fund, awarded to McGill University for the Healthy Brains for Healthy Lives initiative. We also thank the developers of all the applications described with Boutiques.

### Competing interests

The authors declare that they have no competing interests.

---

15 http://www.commontk.org